# Characterization of the probability and information entropy of a process with an exponentially increasing sample space and its application to the Broad Money Supply

Laurence Francis Lacey

Lacey Solutions Ltd, Skerries, County Dublin, Ireland

# Characterization of the probability and information entropy of a process with an exponentially increasing sample space and its application to the Broad Money Supply

## Abstract

There is a random variable (X) with a determined outcome (i.e., X = $x_0$), $p(x_0) = 1$. Consider $x_0$ to have a discrete uniform distribution over the integer interval [1, s], where the size of the sample space (s) = 1, in the initial state, such that $p(x_0) = 1$. What is the probability of $x_0$ and the associated information entropy (H), as s increases exponentially? If the sample space expansion occurs at an exponential rate (rate constant = λ) with time (t) and applying time scaling, such that T = λ x t, gives: $\mathrm{p}(x_0|T) = \exp(-\mathrm{T})$ and $\mathrm{H(T)} = \mathrm{T}$. The characterization has also been extended to include exponential expansion by means of simultaneous, independent processes, as well as the more general multi-exponential case. The methodology was applied to the expansion of the broad money supply of US$ over the period 2001-2019, as a real-world example. At any given time, the information entropy is related to the rate at which the sample space is expanding. In the context of the expansion of the broad money supply, the information entropy could be considered to be related to the "velocity" of the expansion of the money supply.

*Keywords:* statistical methodology, uniform distribution, time series, macroeconomics

# 1. Introduction

The expression for Shannon information entropy [1] is of the form:

$$H = -\sum_{i}^{n} p_i \, log_b \, p_i$$

where $p_i$ is the probability of the i$^{th}$ event occurring out of n possible outcomes, and b, is the base of the logarithm to be used. Common values of b are: 2, Euler's number e, and 10, and the unit of information entropy is shannon (or bit) for b = 2, nat for b = e, and hartley for b = 10. [1] In this paper, information entropy will be used in the following form:

$$H = -\sum_{i}^{n} p_i \, log_e \, p_i$$

The information entropy (H) = 0 for a random variable (X) with a determined outcome (i.e., X = $x_0$), p($x_0$) = 1 [2].

The objective of this paper is to characterise the following statistical process: There is a random variable (X) with a determined outcome (i.e., X = $x_0$), p($x_0$) = 1. Consider $x_0$ to have a discrete uniform distribution over the integer interval [1, s], where the size of the sample space (s) = 1 in the initial state, such that p($x_0$) = 1. What is the probability of $x_0$ and the associated information entropy (H), as s increases exponentially?

## 2. Methods

For a sample space of size, $s_0 > 1$, let $x_0$ have a discrete uniform distribution over the integer interval $[1, s_0]$, giving:

$$\mathrm{p}(x_0) = \frac{1}{s_0}$$

where, $p(x_0)$ is the probability mass function over the integer interval $[1, s_0]$. The cumulative probability of $p(x_0)$ over $[1, s_0]$ is equal to 1. Let $p(x_0|s_0)$ be the <u>cumulative probability</u> of $p(x_0)$ over $[1, s_0]$, i.e.,

$$\mathrm{p}(x_0|s_0) \; = \; 1$$

Double the sample space in which $x_0$ will be uniformly distributed, i.e., $s_1 = 2 \text{ x } s_0$, and partition the new sample space into 2 equally sized, mutually exclusive sample spaces, each of size $s_0$. Now:

$$\mathrm{p}(x_0) = \frac{1}{s_1} = \frac{1}{2 \text{ x } s_0}$$

The cumulative probability of $x_0$, in any <u>one</u> of the two partitions of size $s_0$, is now:

$$\mathrm{p}(x_0|s_1) \; = \; ½$$

Repeating the process a second time, the cumulative probability of $x_0$, in any <u>one</u> of the four partitions, each of size $s_0$, is now:

$$\mathrm{p}(x_0|s_2) \; = \; ¼$$

Repeating the process n times, the cumulative probability, for $x_0$ in any <u>given</u> partition of size $s_0$ is:

$$p(x_0|s_n) = (½)^n$$

This is an exponential process, with the sample space expanding at an exponential rate, giving:

$$p(x_0|s_n) = \exp(-log_e 2 \text{ x } n)$$

The corresponding information entropy of the system is given by:

$$H(n) = -2^n x\ (½)^n\ x\ log_e(½)^n \ = n\ x\ log_e(2)$$

If the expansion occurs at an exponential rate (rate constant = λ) with time (t) and applying time scaling, such that T = λ x t, gives

$$p(x_0|T) = \exp(-T)$$

and,

$$H(T) = T$$

# 3. Results

## 3.1 Mono-exponentially increasing sample space

The attributes of the mono-exponential expansion of the sample space characterized above is shown below in Figures 1 to 3.

Figure 1: Mono-exponential expansion of the sample space, $s_0$ with time (T)

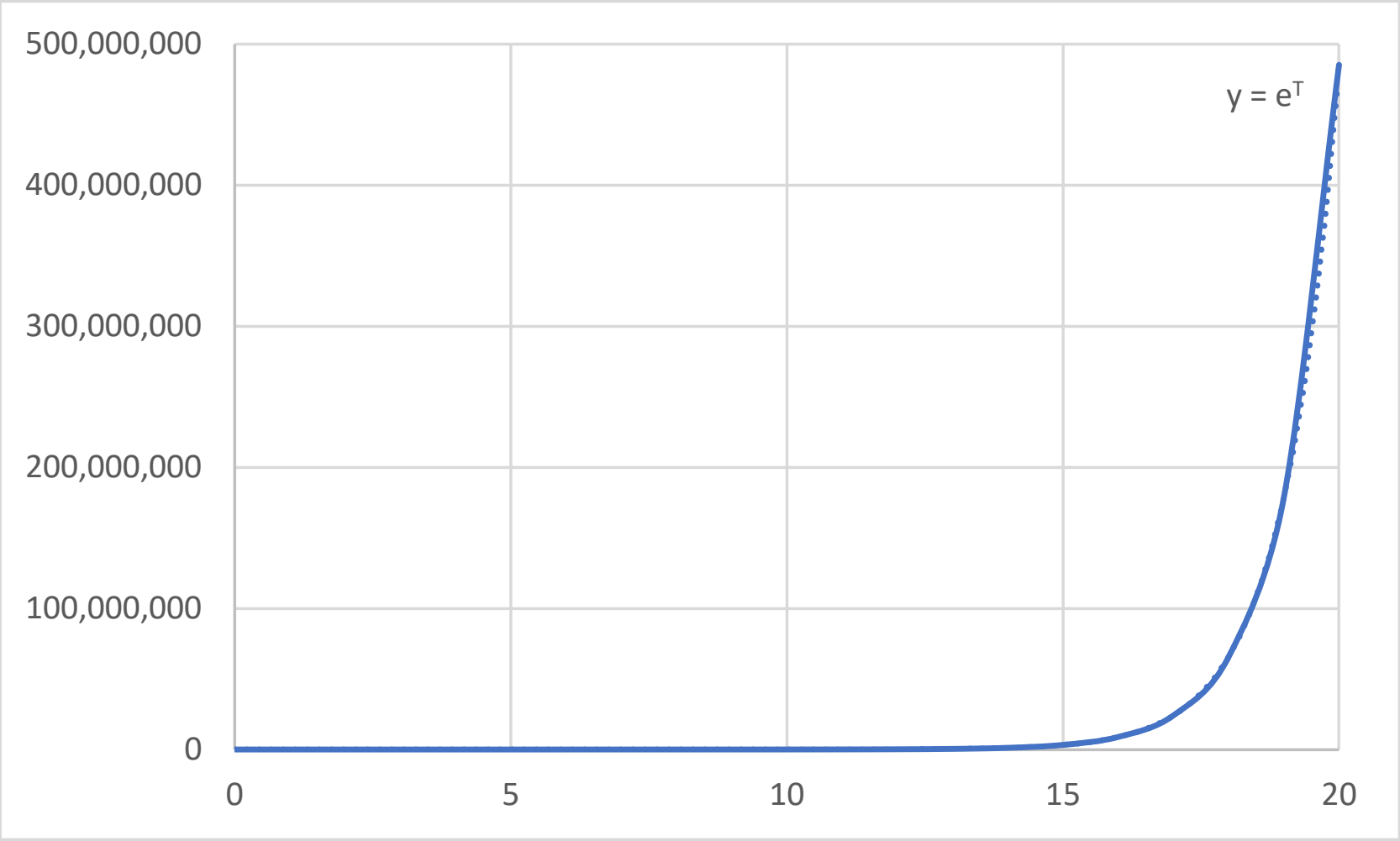

Figure 2: The decrease in the probability with time (T), as a result of the time-dependent expansion of the sample space

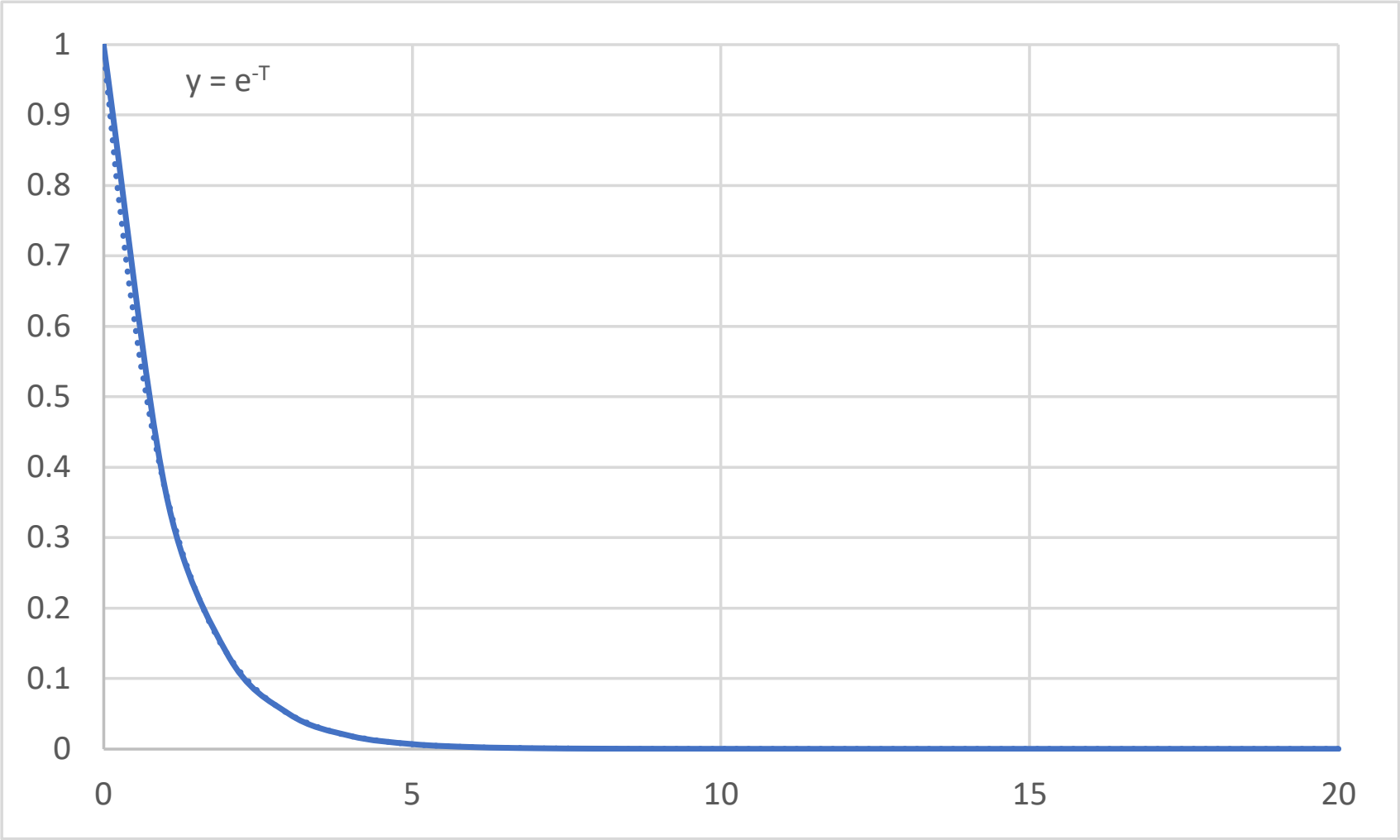

Figure 3: The increase in the information entropy with time (T), as a result of the time-dependent expansion of the sample space

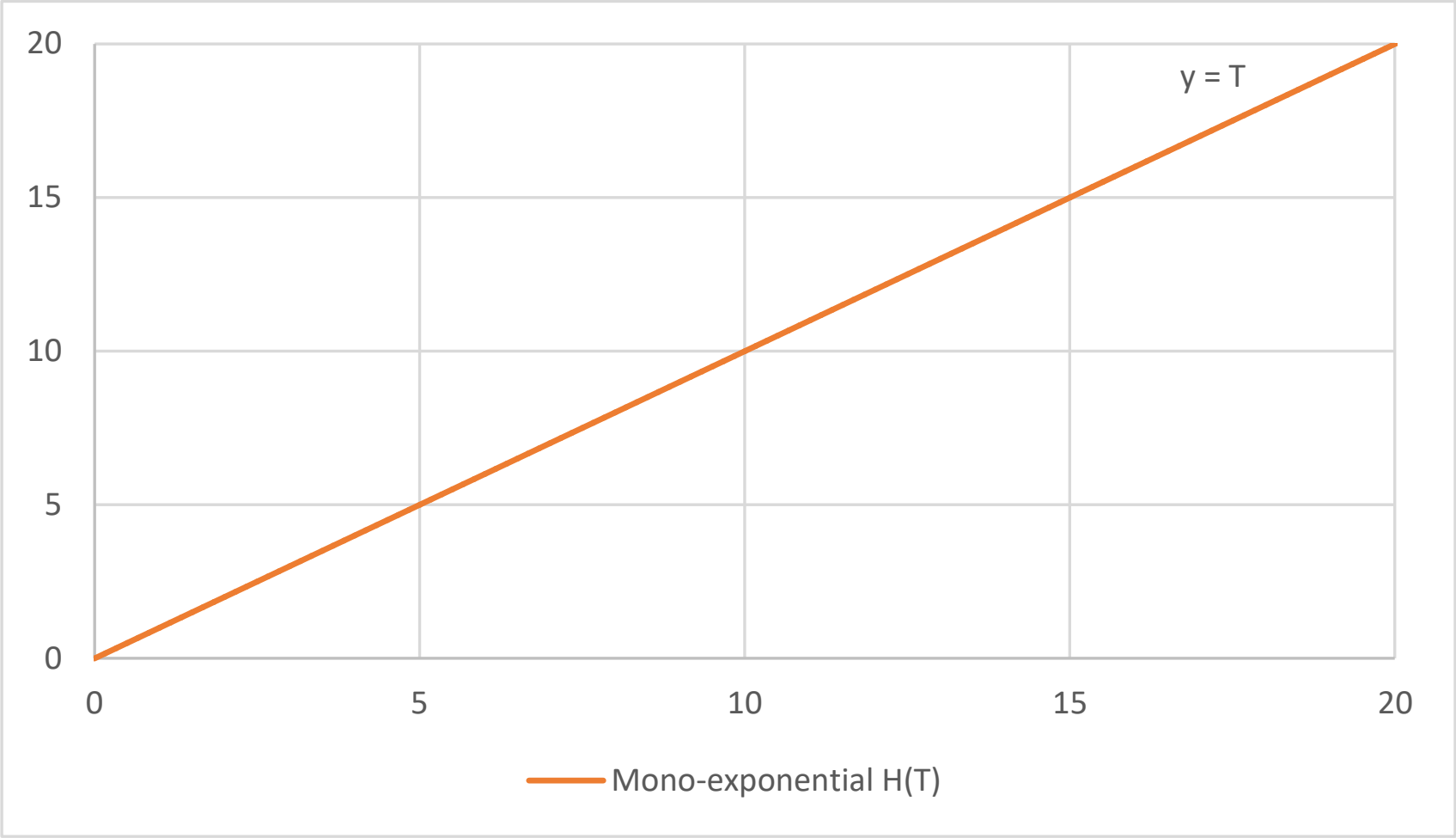

## 3.2 Exponential expansion by means of simultaneous, independent processes

The system's sample space does not have to expand by a single process. It may occur via two or more (n) different processes simultaneously. In general, these processes have different probabilities of occurring, and thus occur at different rates, in parallel. The total exponential rate is given by the sum of $\lambda_1 + \lambda_2 + ..... + \lambda_n$, which can be treated as a new exponential rate constant $\lambda_c$ [3]. In this case, $T' = \lambda_c \times t$.

Comparing an exponential expansion via a single process (rate constant = λ) to that occurring by means of n independent, simultaneous processes (rate constant = $\lambda_c = \lambda_1 + \lambda_2 + ..... + \lambda_n$), the information entropy is the same, if $\lambda = \lambda_c$, with

$$\mathrm{H(T) = T}$$

The attributes of this mono-exponential expansion by mean of three independent simultaneous processes (P1, P2, P3, with $\lambda_1$=0.1; $\lambda_2$=0.3; $\lambda_3$=0.6) is shown below in Figures 4 and 5 below.

In this case,

$$\mathrm{p}(x_0|T) = \mathrm{p}(x_0|P1, T) \text{ x } \mathrm{p}(x_0|P2, T) \; x \; \mathrm{p}(x_0|P3, T)$$

and

$$\mathrm{H(T) = H(T|P1) + H(T|P2) + H(T|P3)}$$

Figure 4: The decrease in the probability with time (T), as a result of the time-dependent expansion of the sample space by mean of three independent simultaneous processes (P1, P2, P3, with $\lambda_1$=0.1; $\lambda_2$=0.3; $\lambda_3$=0.6)

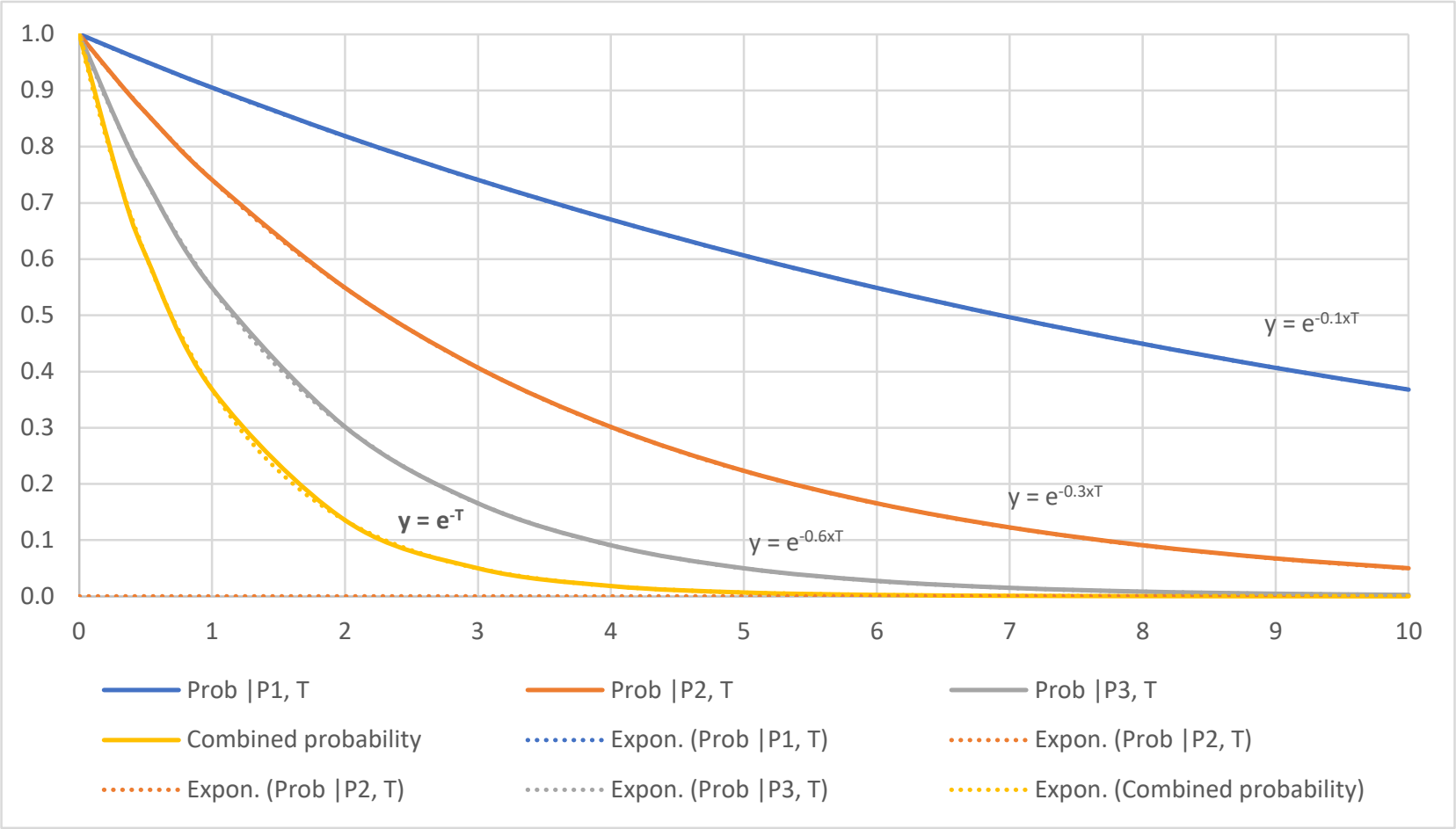

Figure 5: The increase in the information entropy with time (T), as a result of the time-dependent expansion of the sample space by mean of three independent simultaneous processes (P1, P2, P3, with $\lambda_1$=0.1; $\lambda_2$=0.3; $\lambda_3$=0.6)

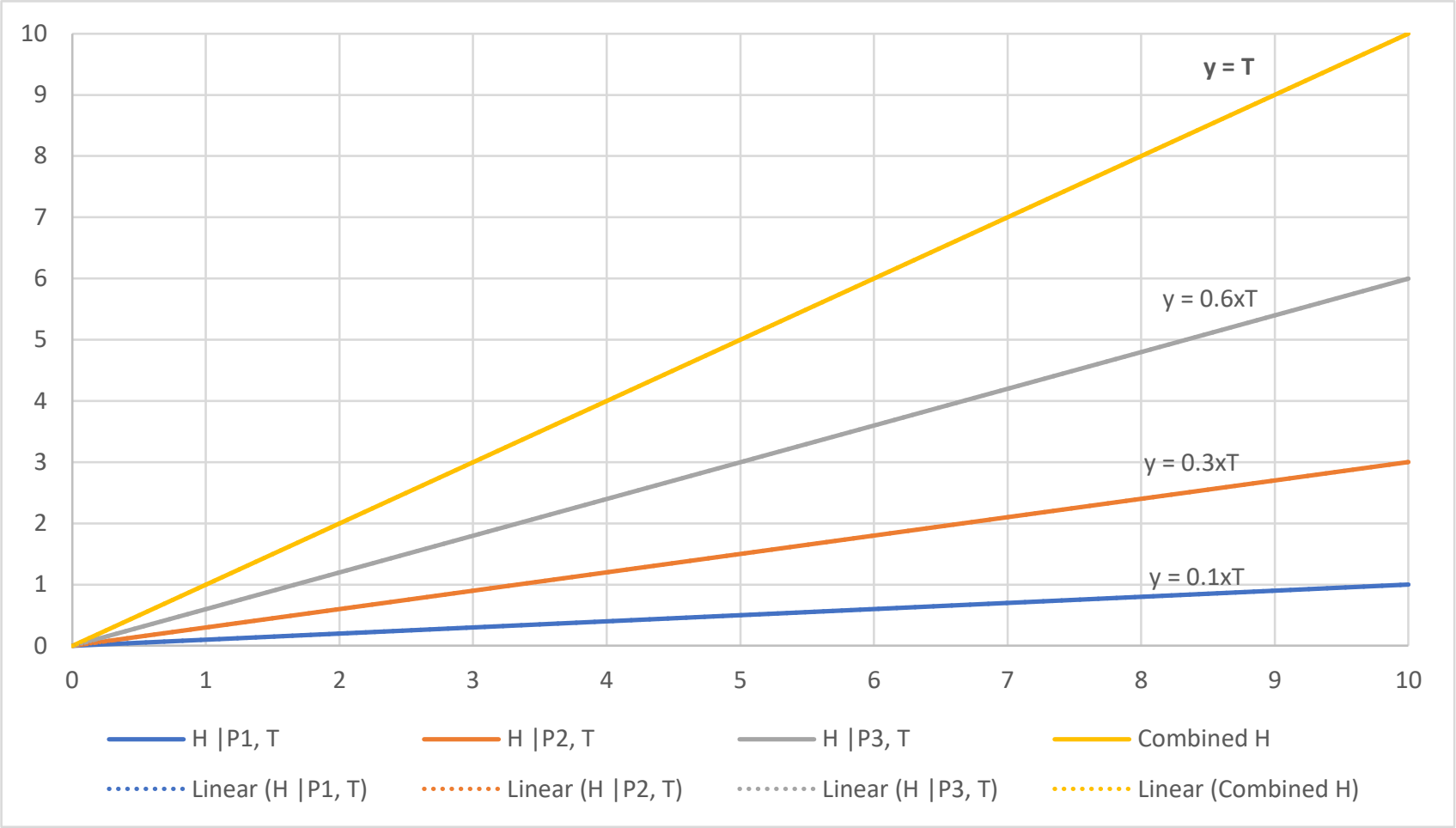

## 3.3 Multi-exponential expansion of the sample space in the more general case

If the sample space, $s_0$, expands multi-exponentially with time:

$$\mathrm{p}(x_0|T) = \sum_{i}^{n} A_i \ \exp(-\mathrm{T} \ \mathrm{x} \ c_i)$$

with,

$$\sum_{i}^{n} A_i = 1$$

where i is the $i^{th}$ term of a multi-exponential function, with n terms, and $c_1 = 1$ and $c_i = \lambda_i/\lambda_1$.

The mean residence time of the sample space (MRT) is given by:

$$MRT = \int_0^\infty \mathrm{T} \ \mathrm{x} \ \mathrm{p}(x_0|T) \ dT \Big/ \int_0^\infty \mathrm{p}(x_0|T) \ dT$$

where,

$$\int_0^\infty \mathrm{p}(x_0|T) \ dT = \sum_{i}^{n} A_i \, / \, c_i$$

$$\int_0^\infty \mathrm{T} \ \mathrm{x} \ \mathrm{p}(x_0|T) \ dT = \sum_{i}^{n} A_i \, / ( \, c_i \,)^2$$

The value of H(T) for this multi-exponential expansion of the sample space is:

$$\mathrm{H(T)} = -\, log_e \left( \sum_i^n A_i \ \exp(-\mathrm{T} \times c_i) \right)$$

When T is large,

$$\mathrm{H(T)} = (\mathrm{T} \times c_n) - log_e \, (A_n)$$

Normalized H(T) is obtained by dividing H(T) by H($T_{max}$), where $T_{max}$ is the (arbitrary) maximum value of T.

$$\text{Normalized H(T)} = \mathrm{H(T)} \,/\, \mathrm{H(T_{max})}$$

An example is given below, in which the sample space expands multi-exponentially with time, with four distinct exponential components, the parameterization of which is given in Table 1. The multi-exponential expansion of the sample space with time is shown as a semi-log plot in Figure 6. The corresponding multi-exponential decline in $\mathrm{p}(x_0|T)$ is shown in Figure 7, and the increase in H(T) with time, is shown in Figure 8.

Table 1: Multi-exponentially expansion of the sample space with time, with four separate exponential components

| **n = 1, 2, 3, 4 exponential components** | $A_n$ | $c_n$ |
|---|---|---|
| 1 | 0.4 | 1.0 |
| 2 | 0.3 | 0.1 |
| 3 | 0.2 | 0.01 |
| 4 | 0.1 | 0.001 |

Figure 6: Multi-exponential expansion of the sample space with time (T) (semi-log plot)

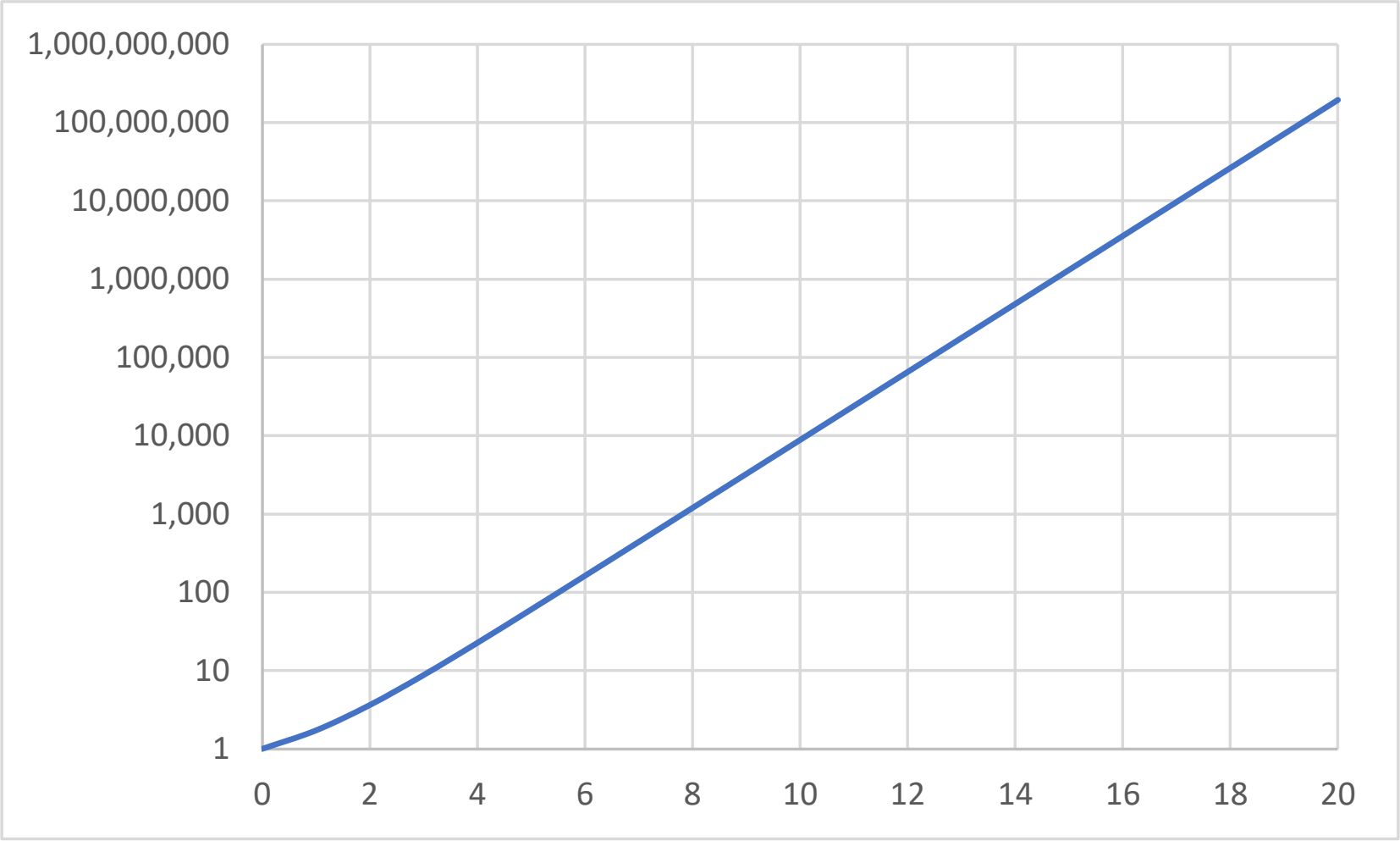

Figure 7: The decrease in the probability with time (T), as a result of the time-dependent multi-exponential expansion of the sample space, in which, $\mathrm{p}(x_0|T) = \sum_i^4 A_i \exp(-\mathrm{T} \times c_i)$

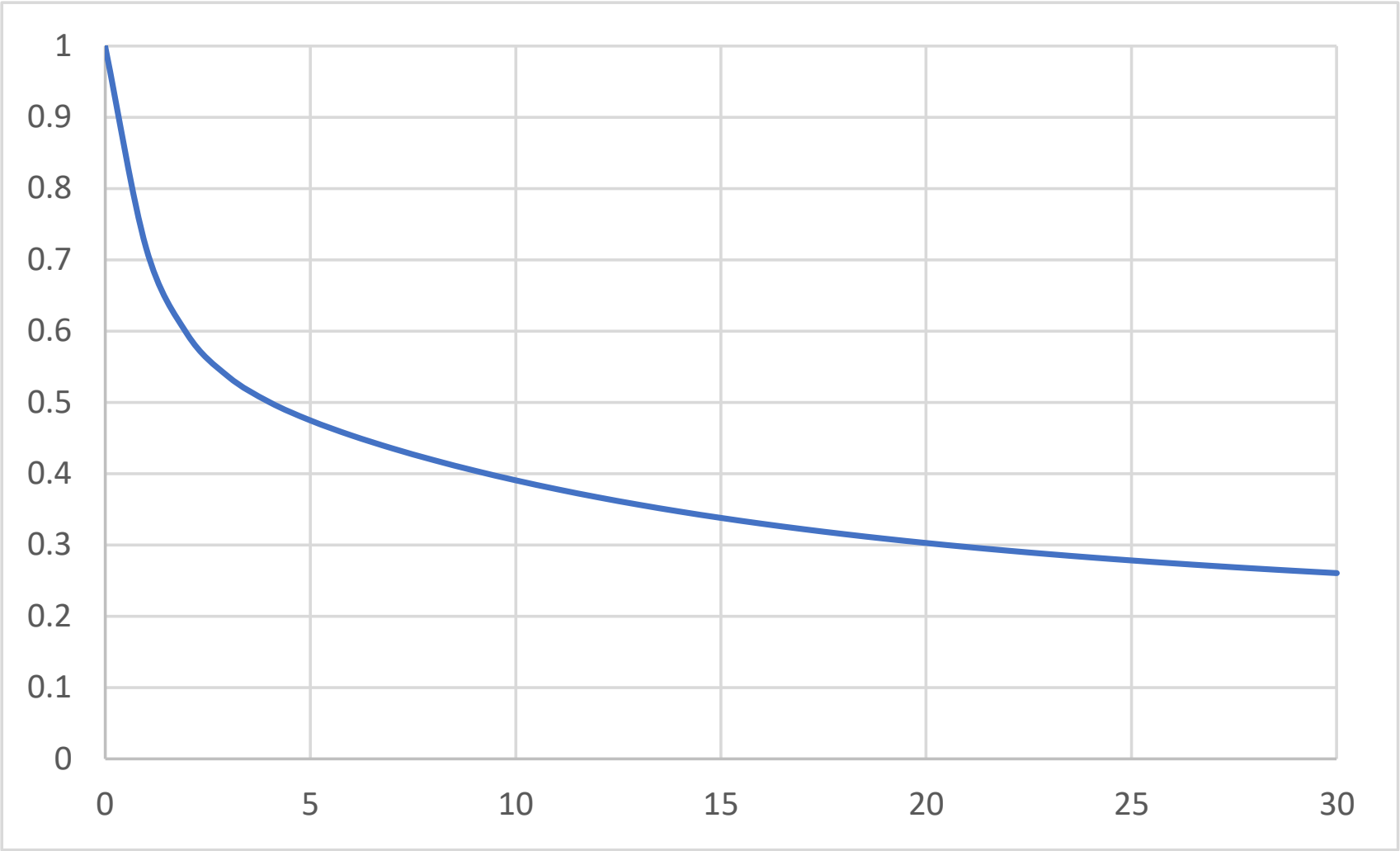

Figure 8: The increase in the information entropy H(T) with time (T), as a result of the time-dependent multi-exponential expansion of the sample space, in which, $\mathrm{H(T)} = -\log_e \left(\sum_i^4 A_i \exp(-\mathrm{T} \times c_i)\right)$

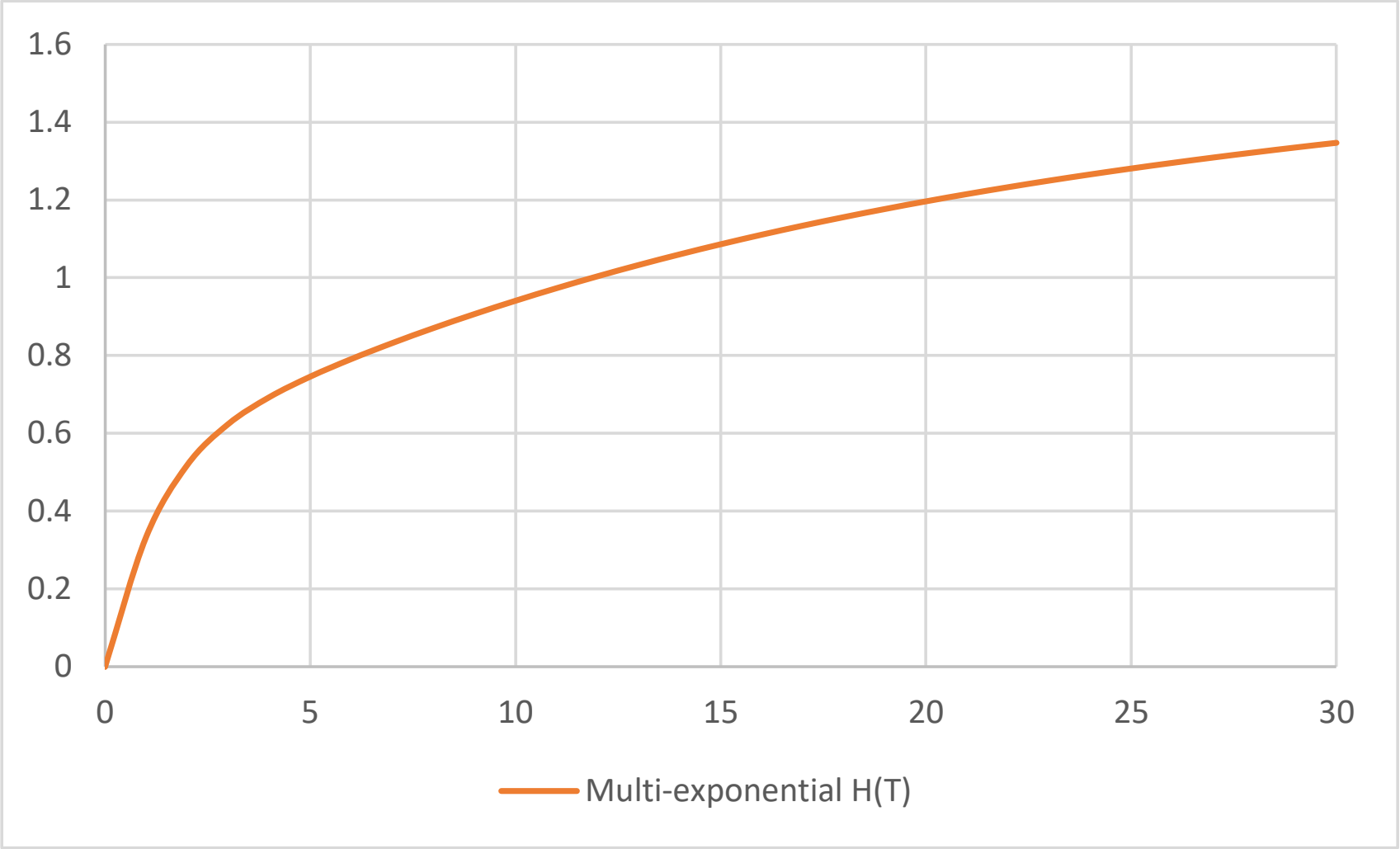

The contribution of each of the 4 components of the multi-exponential expansion of the sample space, is shown for decline in the probability, $\mathrm{p}(x_0|T)$, in Figure 9 and for the increase in the information entropy, H(T), in Figure 10.

Figure 9: Multi-exponentially expansion of the sample space with time, with four separate exponential components, in which the contribution of each component in shown in relation to $\mathrm{p}(x_0|T)$, as T increases

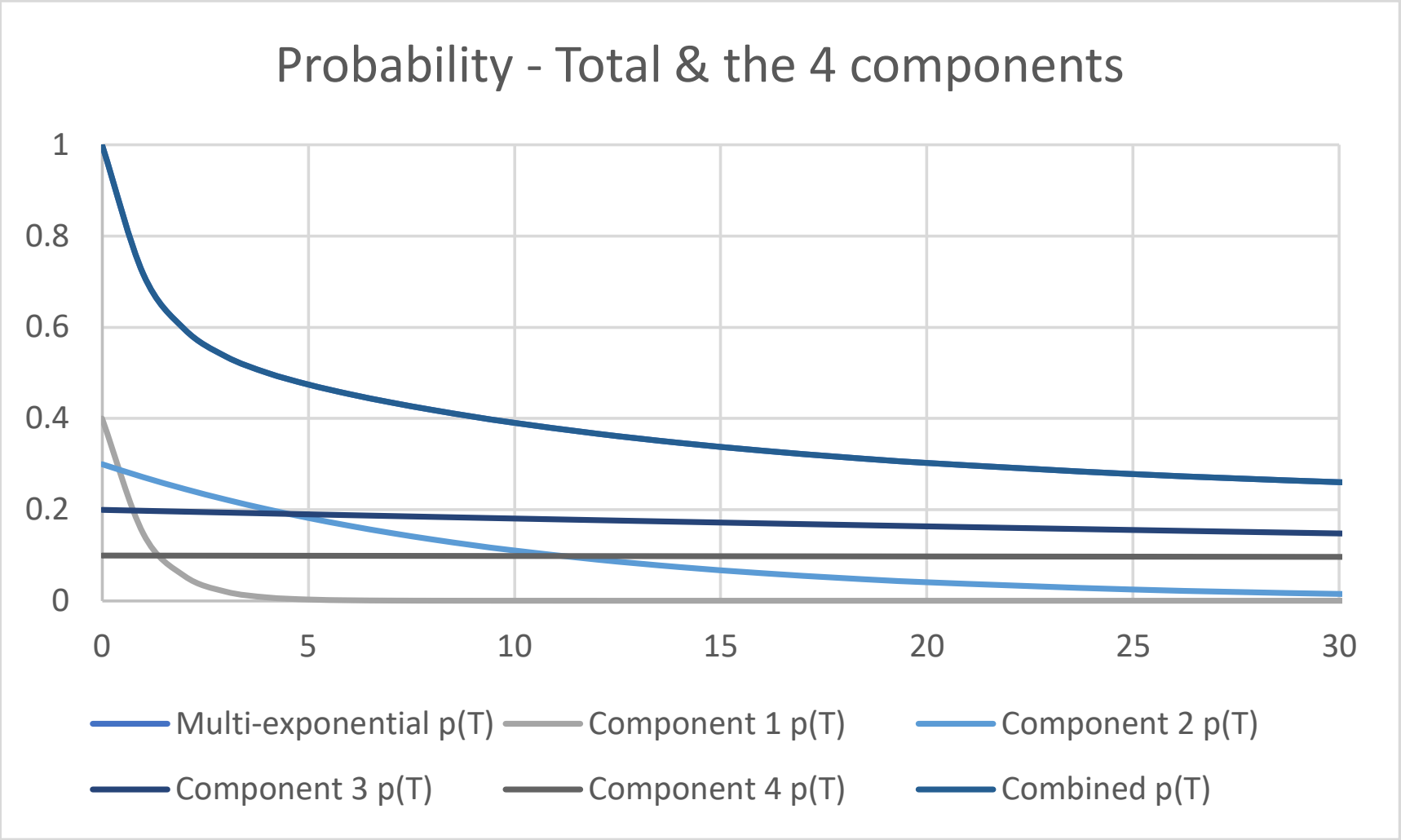

Figure 10: Multi-exponentially expansion of the sample space with time, with four separate exponential components, in which the contribution of each component in shown in relation to total information entropy (H(t)), as T increases

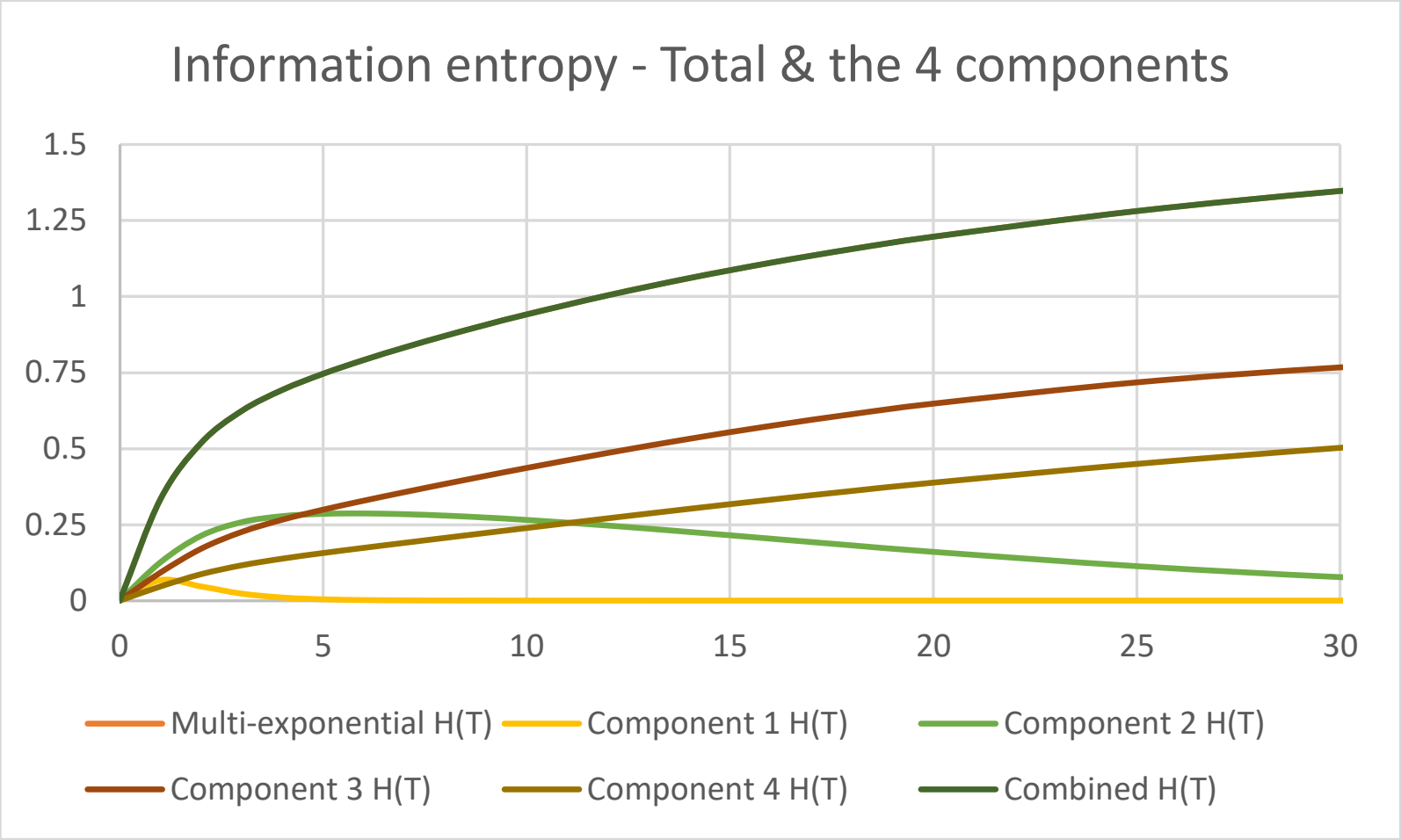

When T becomes large, the long-term linear increase of H(T) with T is due to the exponential component with the smallest rate-constant, $c_4$ = 0.001 (Figure 11, Table 1). The normalized H(T) is given in Figure 12, in which $T_{max}$ is arbitrarily set to T=1,000.

Figure 11: Multi-exponentially expansion of the sample space with time, with four separate exponential components, in which the contribution of each component in shown in relation to total information entropy (H(t)), as T becomes large

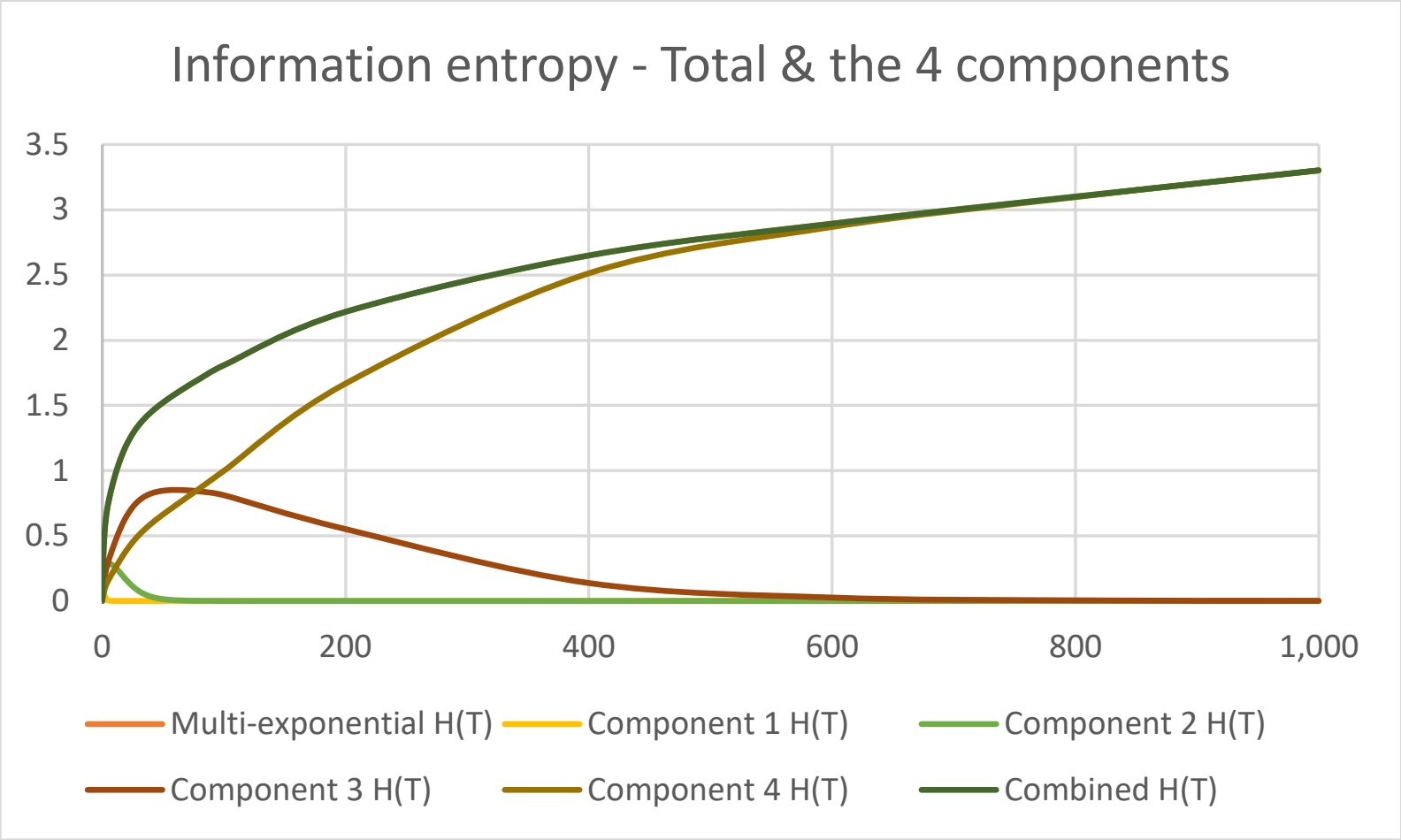

Figure 12: The normalized H(T) for the multi-exponential expansion of the sample space $\left(\frac{H(T)}{H(T_{max})}\right)$, in which $T_{max}$ is arbitrarily set to T=1,000

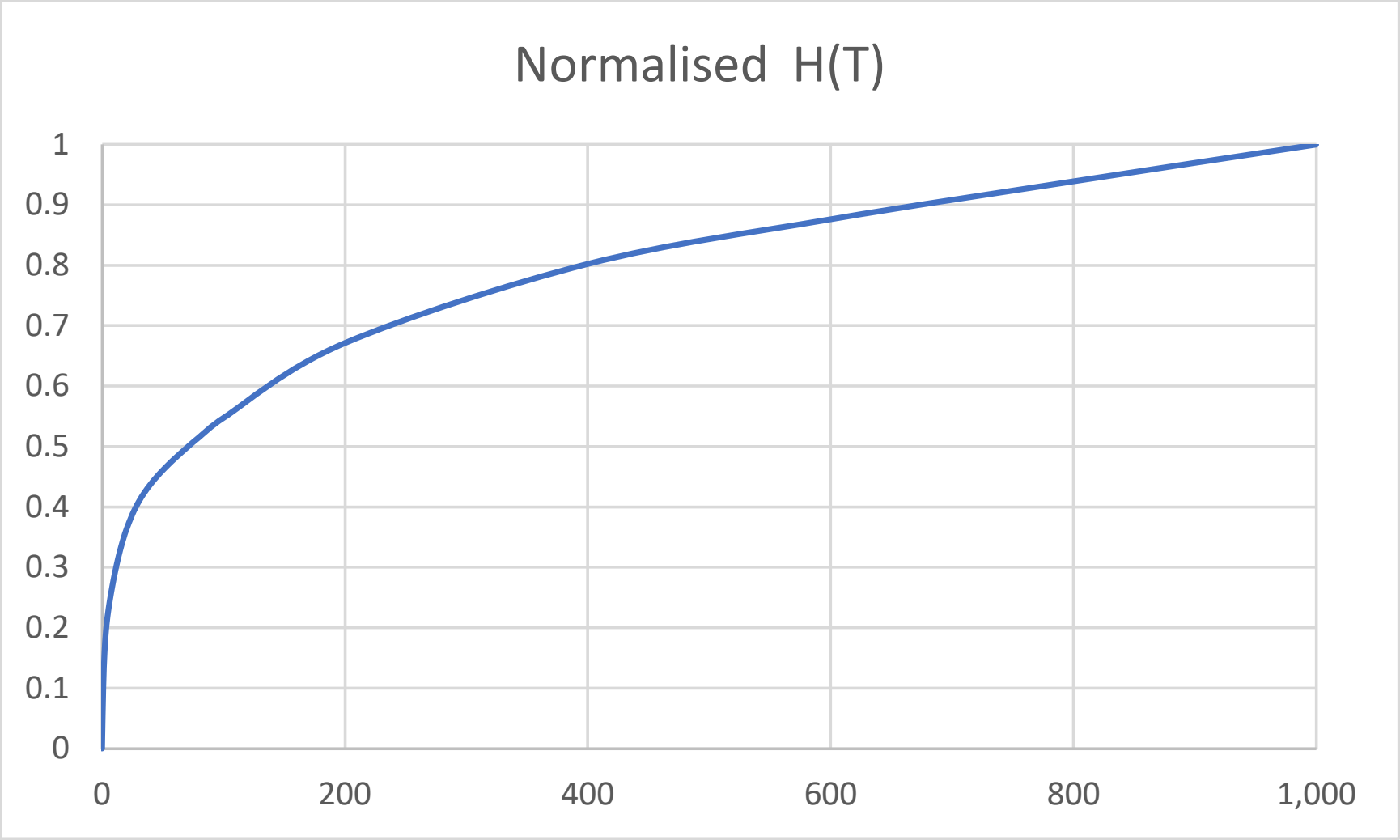

This completes the characterization of the probability and information entropy of a statistical process with an exponentially increasing sample space.

### 3.4 Non-exponential expansion of the sample space

The methodology can easily be adapted to other types of sample space expansion, e.g., power function expansion. However, that is outside the scope of this paper.

### 3.5 Exponential contraction of the sample space

The methodology can also easily be adapted to contraction of the sample space. This is characterized for a mono-exponential contraction, involving three simultaneous, independent processes in the Appendix.

### 3.6 Real world application – expansion of the broad money supply

The following will provide an example of applying the methodology outlined above to the expansion of the broad money supply of US$.

Broad money includes both notes and coins, but also other forms of money, which can easily be converted into cash [4]. The expansion of the broad money supply of US$ from 2001 to 2019 was obtained from the International Monetary Fund (IMF) data on monetary and financial statistics [5].

To apply the methodology outlined above, we set the size of the initial sample space ($s_0$) equal to the broad money supply of US$ in circulation in 2001. Also, we set 2001 as t=0, i.e., 2001 is set as the reference sample space for the time series. The broad money supply of US$ from 2001 to 2019 (time, 0 to 18 years) is plotted in Figure 13. As can been seen, the expansion of the money supply can be well described using an exponential function, in which,

$$s_0 = 7.5805\ trillion\ US\$$$

The coefficient of determination ($R^2$) of the fit is approx. 98% for the 19-year time series. Over this period, the exponential rate-constant (λ) = 0.0555. This is equivalent to a mean annual rate of increase in the broad money supply of US$ over this time period of 5.7%.

Figure 13: Mono-exponential expansion of the broad money supply of US$ from 2001 to 2019 (t = 0 to 18 years) (in trillions of US$)

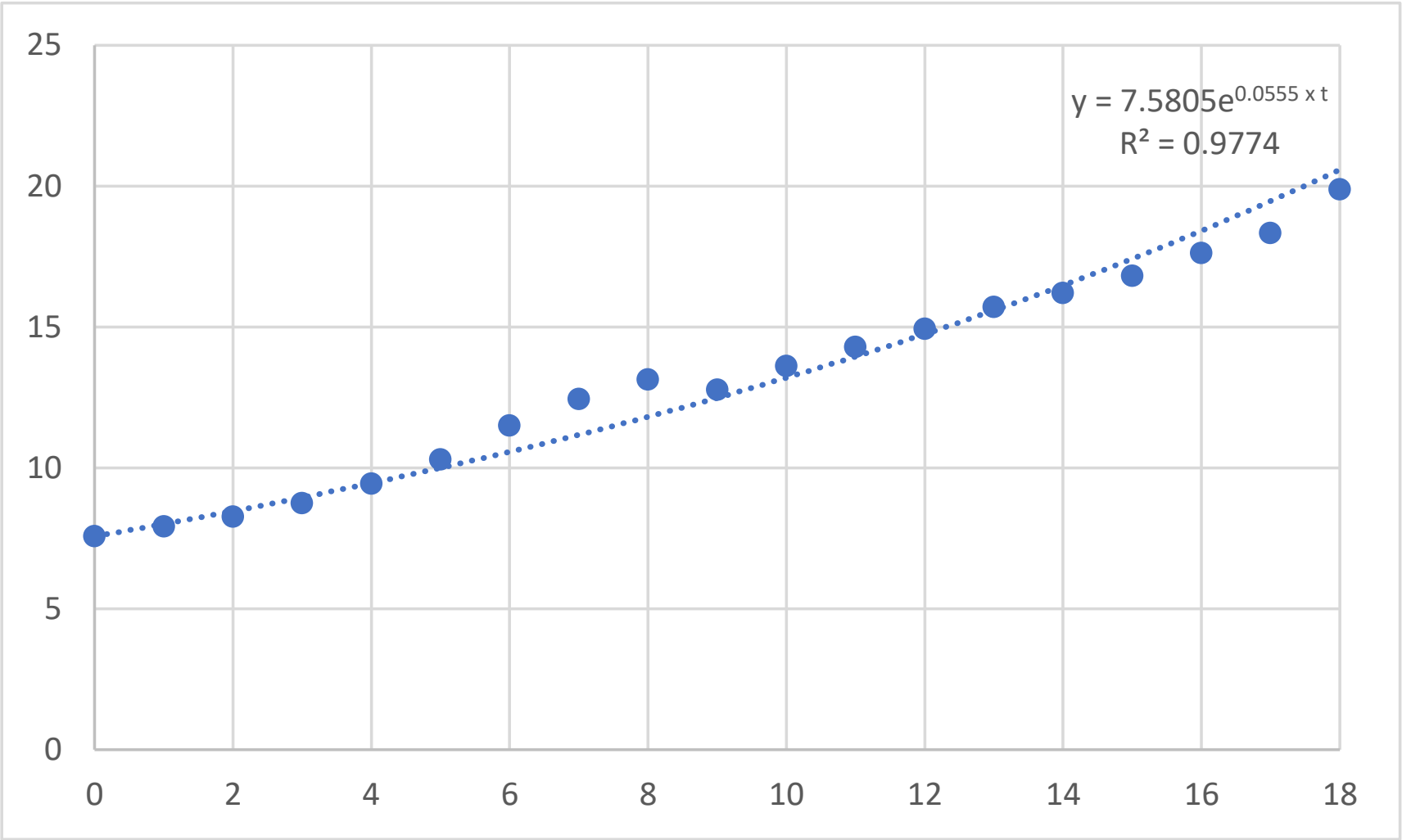

The probability $\mathrm{p}(x_0|t)$ over the time period 2001 to 2019 (t = 0 to 18 year) will be:

$$\mathrm{p}(x_0|t) = 7.5805\ x\ \exp\left(-0.0555\ x\ t\right)$$

The corresponding information entropy of the broad money supply of US$ over the same time period is:

$$H(t) = 0.0555 \; x \; t$$

which is as shown in Figure 14, with H(t=0) = 0, because 2001 was set as the reference sample space for the time series.

Figure 14: The increase in the information entropy of the broad money supply of US$ over the period 2001 to 2019 (t = 0 to 18 years)

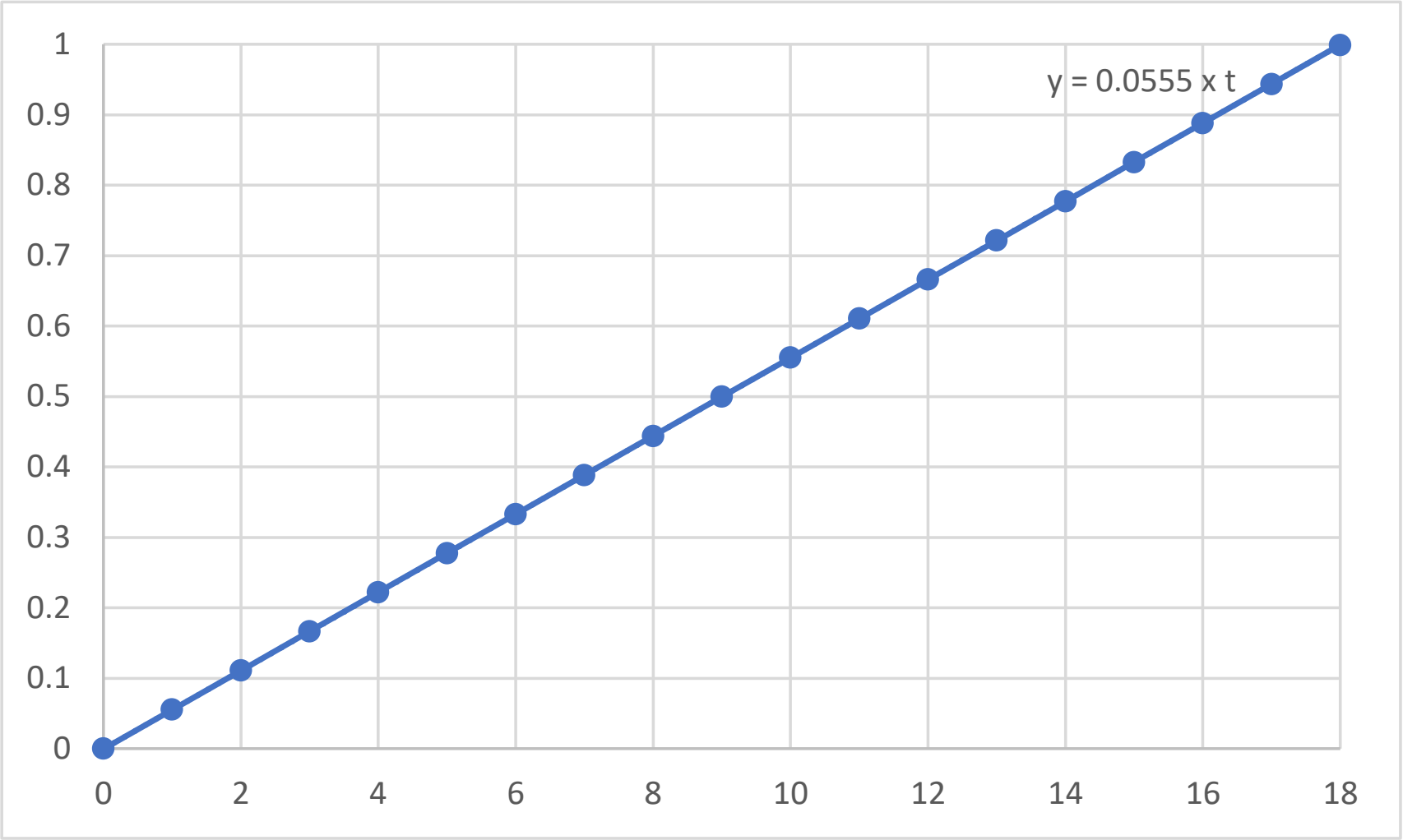

## 4. Discussion

The probability and information entropy of a process with an exponentially increasing sample space has been characterized in the case of (1) a mono-exponential expansion of the sample space; (2) exponential expansion by means of simultaneous, independent processes; (3)

the more general multi-exponential expansion of the sample space. In addition, the methodology was shown to be easily adaptable to mono-exponential contraction of the sample space, through one or more processes (see Appendix).

The methodology can easily be adapted to other types of sample space expansion, e.g., power function expansion. However, that was outside the scope of this paper and would need a subsequent paper to demonstrate this.

The methodology was applied to the expansion of the broad money supply, using the US$ as a real-world example. An in-depth analysis of the broad money supply using the methodology was outside the scope of this paper. The exercise was simply done to demonstrate that the analysis of the time-series of the broad money supply could be undertaken using the methodology. This resulted in a "metric" not currently used in the macro-economic literature, the information entropy of the broad money supply. What is it?

The information entropy of an exponentially expanding sample space has no units, because λ has the units, $time^{-1}$, so that time x λ has no units ($time^{-1}$ x time). However, at any given time, the information entropy is related to the rate at which the sample space is expanding. In the context of the expansion of the broad money supply, the information entropy could be considered to be related to the "velocity" of the expansion of the money supply. This is completely different to the existing economic term, the "velocity of money" [6].

It was outside the scope of the current paper to investigate how the information entropy of the broad money supply could be used to answer questions in macro-economics. This would need to be the subject of a stand-alone paper.

It is possible that other real-world examples, involving expanding or contracting processes may be usefully analysed using the methodology described in this paper.

## 5. Conclusion

The probability and information entropy of a process with an exponentially increasing sample space has been characterized. The methodology can be easily adaptable to the exponential contraction of the sample space. A key outcome of the methodology is the information entropy of the process under investigation. The methodology was applied to the expansion of the broad money supply of the US$, as an example. It is possible that other real-world examples, involving expanding or contracting processes may be usefully analysed using the methodology described in this paper.

## Supplementary materials

Supplementary materials relating to a statistical process with an exponentially decreasing sample space are given in the appendix.

## Acknowledgements

The author gratefully acknowledges those academic and non-academic people who kindly provided feedback on earlier drafts of this paper. No financial support was received for any aspect of this research.

# APPENDIX: Characterization of the probability and information entropy of a statistical process with an exponentially decreasing sample space

The information entropy (H) = 0 for a random variable (X) with a determined outcome (i.e., $X = x_0$), $p(x_0) = 1$ [2]. Consider $x_0$ to have a "discrete uniform distribution" over the integer interval [1, s], where initially the sample space (s) = 1, such that $p(x_0) = 1$

When $s_0 > 1$, $x_0$ has a discrete uniform distribution over the integer interval $[1, s_0]$, giving:

$$\mathrm{p}(x_0|s_0) = \frac{1}{s_0}$$

where, $\mathrm{p}(x_0|s_0)$ is the probability mass function over the integer interval $[1, s_0]$.

Let us half the size of the sample space, such that it is now, $s_1 = s_0/2$, giving:

$$\mathrm{p}(x_0|s_1) = \frac{2}{s_0}$$

Halving the size of the sample space a second time, such that it is now, $s_2 = s_0/4$, gives:

$$\mathrm{p}(x_0|s_2) = \frac{4}{s_0}$$

Repeating the process n times, gives $s_n = s_0/(2^n)$, and

$$p(x_0|s_n) = \frac{2^n}{s_0} = \exp(log_e 2\ x\ n)$$

and

$$n = \ log_e 2/log_e s_0$$

The process can only occur n times in order to comply with the condition that:

$$p(x_0|s_n)\ \leq 1$$

The corresponding information entropy (H) of the process is given by:

$$H(n) = -\ n\ x\ log_e(2)$$

When the contraction process occurs with time (t):

$$p(x_0|t) = \exp(log_e 2\ x\ t)$$

$$H(t) = -\ t\ x\ log_e(2)$$

And the process can continue only until the following condition is reached:

$$t \leq\ log_e 2/log_e s_0$$

An example is provided below. Let $s_0$ = [1, 1,000], and let the process occur with time (t) such that at t=0:

$$p(x_0|s_0) = \frac{1}{1{,}000} = 0.001$$

In this case, the process continues until:

$$t = \frac{log_e 2}{log_e 1{,}000} = \text{approx. } 10$$

at which time:

$$p(x_0|t = \ approx\ 10) = 1$$

The contraction of the sample space, $s_0$ is given in Appendix Figure 1. The increase in the probability with contraction of the sample space is given in Appendix Figure 2, and the information entropy (H) of the contraction process is given in Appendix Figure 3.

Appendix Figure 1: Mono-exponential contraction of the sample space, $s_0$ with time (t)

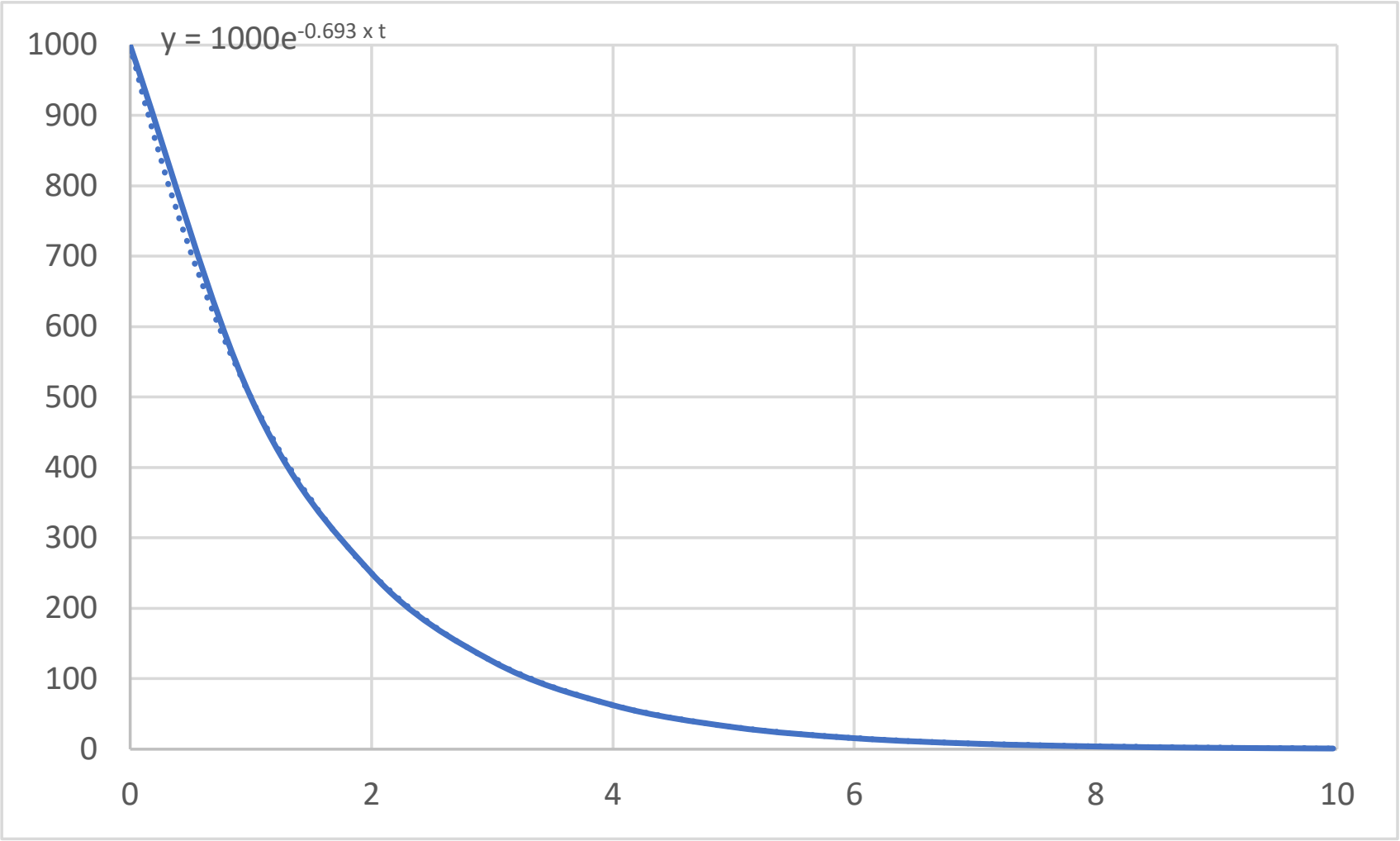

Appendix Figure 2: The increase in the probability with time (t), as a result of the time-dependent contraction of the sample space, until t = approx. 10.

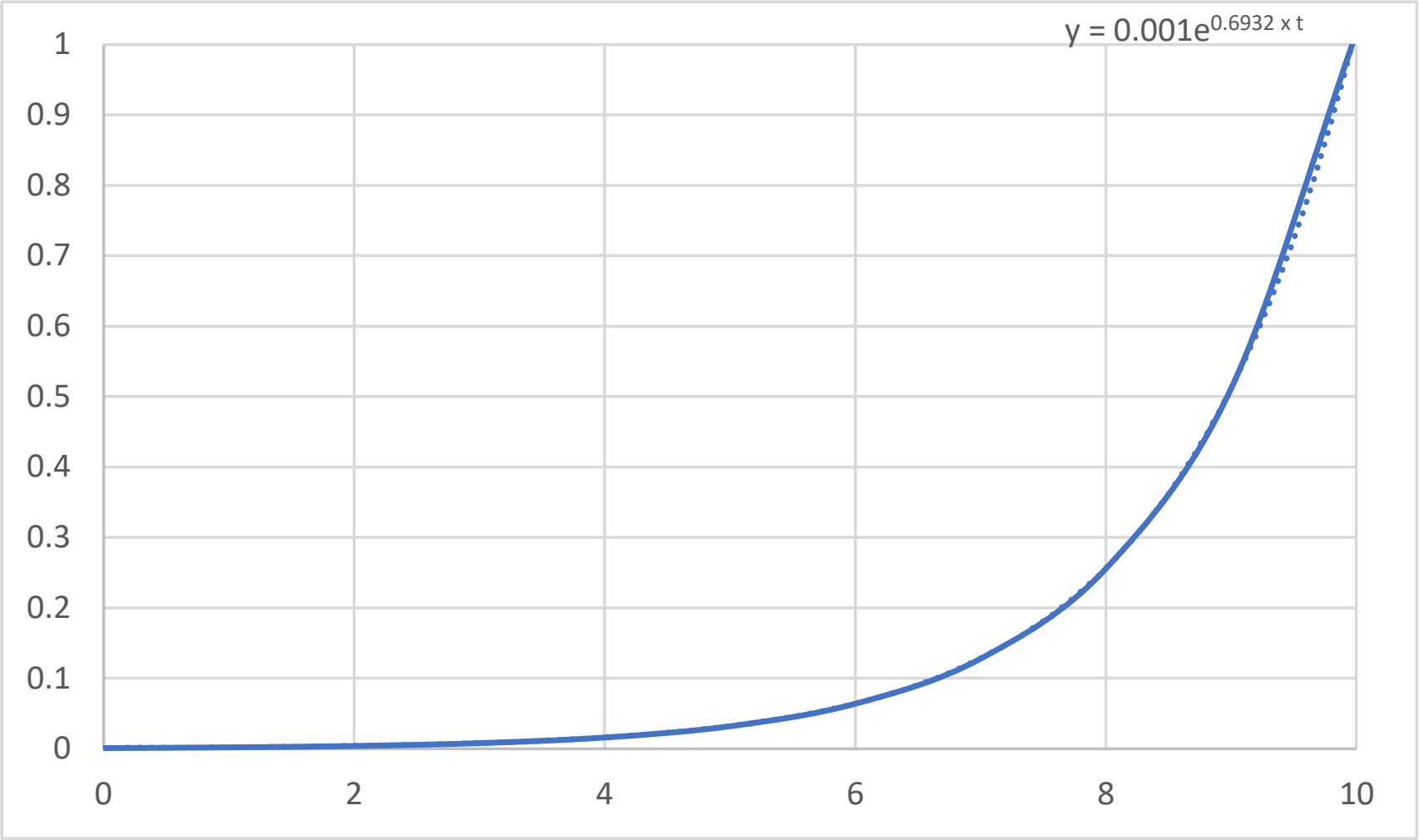

Appendix Figure 3: The decrease in the information entropy with time (t), as a result of the time-dependent contraction of the sample space, until t = approx. 10.9

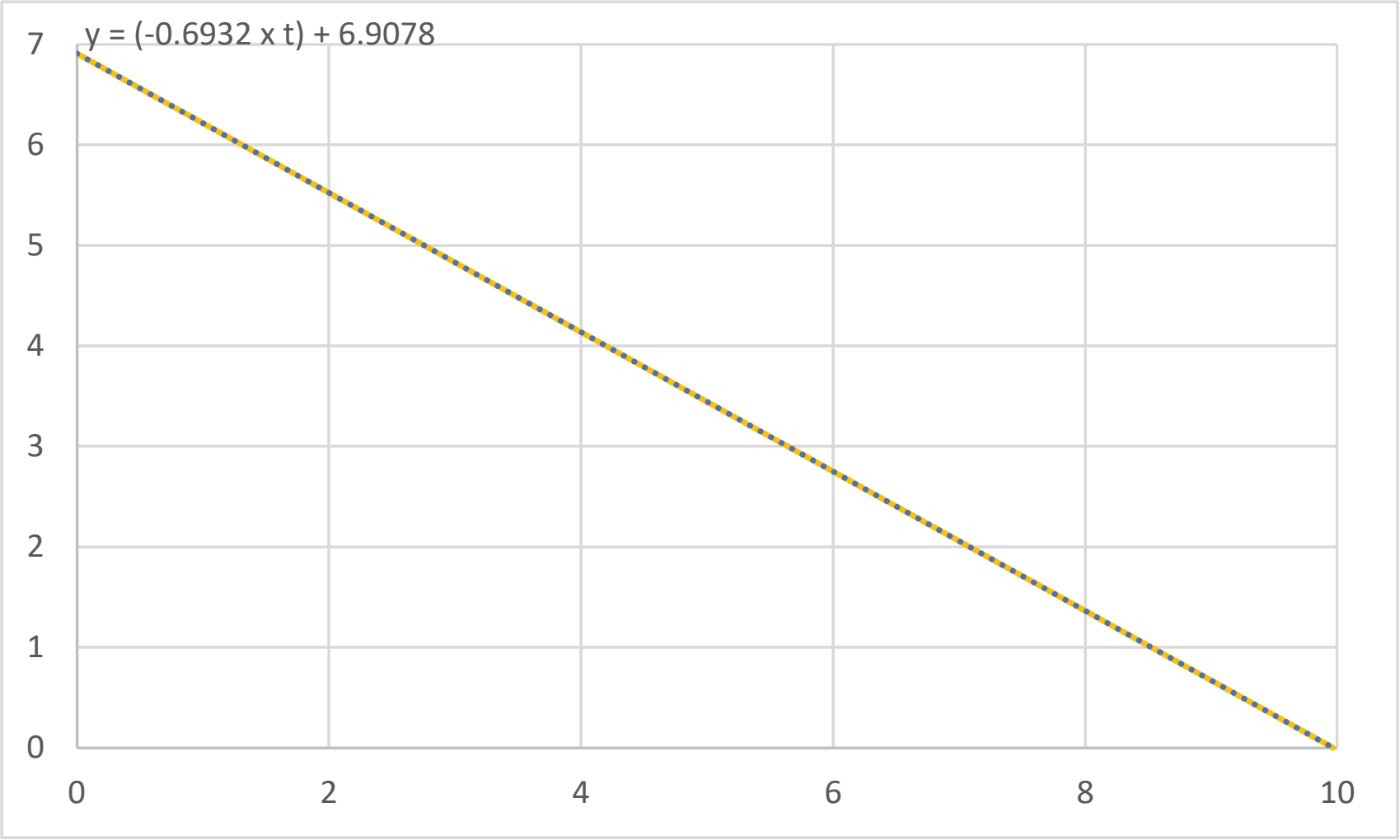